\documentclass[twocolumn,pre,floatfix,superscriptaddress]{revtex4}
\usepackage{graphicx}
\usepackage{times}
\usepackage{color}
\usepackage{amsmath}
\usepackage{epstopdf}
\def\D{{\rm d}}
\def\E{{\rm e}}
\def\I{{\rm i \,}}
\begin{document}

\title{Metric theory of nematoelastic shells}
\author{L.~M. Pismen}
\affiliation{Department of Chemical Engineering, Technion -- Israel Institute of Technology, Haifa 32000, Israel}

\begin{abstract}
We consider three-dimensional reshaping of a thin nematoelastic film upon nematic-isotropic transition in the field of a charge one topological defect, leading to either cone or anticone ($d$-cone) shells. The analysis is based on the relation between the shell metric and the tensor order parameter under the assumption of no elastic deformation and volume change. The shape of the shell can be modified by doping, creating cones with curved generatrices. Anticones necessarily have an even number of radial creases. The curvature singularity at the apex is resolved due to decay of the nematic order parameter at the defect core. 

\end{abstract}
\pacs{83.80.Xz, 83.10.Ff, 83.10.Bb} 
 
 \maketitle

Liquid crystal elastomers and glasses \cite{Warner}, made of cross-linked polymeric chains with embedded mesogenic structures, combine orientational properties of liquid crystals with shear strength of solids. Their specific feature is a strong coupling between the director orientation and mechanical deformations, which can be controlled by the various physical and chemical agents, and enables multiple applications to the construction of artificial muscles \cite{degennes97}, biomimetic actuators \cite{YT11}, and artificial swimmers \cite{swim}. When the material undergoes a phase transition from the isotropic to nematic state, it strongly elongates along the director and, accordingly, shrinks in the normal direction to preserve its volume; the opposite effect takes place as result of the reverse transition. The magnitude of elongation can be regulated by changing the concentration of a mesogenic component, which may be caused by radiation-induced isomerization \cite{Samitsu}.

The reshaping effect in thin films can be exploited \cite{Warner12} to bend a flat sheet into a curved shell. Unlike bending isotropic materials \cite{cerda,witten},  this may create shells with non-zero Gaussian in the absence of shear or compression. This can be practically realized in a most efficient way by preparing in the liquid state a nematic alignment field, which becomes frozen after polymerization. The sheet would bend when the temperature or concentration of a non-mesogenic component is lifted above the nematic-isotropic transition point (NIT) and the material becomes isotropic. The problem studied in this communication is the same as in Ref.~\cite{Warner12} but both the method and results are different. Our analysis is based on a metric transformation of a thin film following NIT under the assumption of no elastic deformation and volume change. Unlike Ref.~\cite{Warner12}, we find that transformation involving expanding circumference and shortening radii in the field of a topological defect leads to crumpled rather than smooth shells. We also explore how the curvature singularity is resolved due to a natural decay of the nematic order parameter at the defect location.

%
We consider a thin film with preferential tangential orientation of the nematic director on both upper and lower boundaries. Under these conditions, the material can be described by the 2D nematic order parameter, which we present as a traceless symmetric tensor $Q_{\alpha\beta}$ dependent on 2D coordinates. We will use two sets of coordinates: 3D coordinates $x^i$ in Euclidean space, which reduce to 2D coordinates $x^\alpha$ on a flat sheet, and 2D coordinates $\xi^\alpha$ on a curved shell. The 2D metric induced on the shell is $\gamma_{\alpha\beta}=g_{ij}x^i_{,\alpha}x^j_{,\beta}$, where $g_{ij}$ is a Euclidean 3D metric; Latin indices run from 1 to 3 and Greek, from 1 to 2; commas denote partial derivatives and summation over paired upper and lower indices is presumed throughout.

Under the assumption of negligible tangential elastic deformations, transition from the nematic to the isotropic state causes a shrinkage along the nematic director, so that an infinitesimal interval is transformed as 
\begin{equation}
\D\xi^\alpha=N^{-1/2}\left(\delta^\alpha_\beta - a Q^\alpha{}_\beta \right)\D x^\beta,  
\label{eq:xxi}
\end{equation}
where the metric factor $a$ quantifies the length change, $N$ is a normalization factor, and $\delta^\alpha_\beta$ is the Kronecker delta; the indices are lowered and lifted with the help of the flat metric $g_{\alpha\beta}$ and its inverse. Since the interval can be expressed as $\D s^2=\gamma_{\alpha\beta} \D\xi^\alpha\D\xi^\beta = g_{\alpha\beta} \D x^\alpha\D x^\beta$, the respective metric tensors are related as 
\begin{equation}
g_{\alpha\beta}= N^{-1}(\gamma_{\alpha\beta} - 2a Q_{\alpha\beta}+ a^2 Q^\gamma{}_\alpha Q_{\gamma\beta}) .
 \label{eq:gg}
\end{equation}
Since the director is frozen in the material, the metric transformation should be understood in the Lagrangian sense as applicable to a point shifting in the course of the deformation, so that the metric $\gamma_{\alpha\beta}$ computed at any point of the flat sheet is translated to the image of this point on the bent shell.  

The tensor $Q_{\alpha\beta}$ can be expressed through the unit vector $n_\alpha$ as
$Q_{\alpha\beta} = S \left( n_\alpha n_\beta - \textstyle{\frac 12}g_{\alpha\beta} \right)$,
where $S$ is the scalar order parameter dependent on the deviation from NIT. In Cartesian coordinates,
\begin{equation}
\mathbf{Q }= \frac{S}{2} \left(\begin{array}{cc}\cos 2\theta & \sin 2\theta \\ 
 \sin 2\theta & -\cos 2\theta \end{array} \right) \equiv 
  \left(\begin{array}{cc}p & q \\ q & -p \end{array} \right),
\label{eq:pq} \end{equation} 
where $\theta$ is the director orientation angle. The parameter $S$ can be normalized by incorporating this deviation in the metric factor $a$, which can be made variable by localized doping. We will impose the normalization condition $Q^\alpha{}_\beta Q^\beta{}_\alpha=2$, leading to $S=2$. Using Eq.~\eqref{eq:pq} in Eq.~\eqref{eq:gg} one can compute the normalization factor $N=1-a^2$ ensuring the conserved area condition $g \equiv \det g_{\alpha\beta} =\det \gamma_{\alpha\beta} \equiv \gamma$ for an incompressible material; this result is independent of a choice of either metric. It is easy to see that when the axis $x^1$ is oriented along the director, so that $p=1, \: q=0$, the scale changes along and across the director are expressed through the extension coefficient $\ell$:
\begin{equation}
\D\xi^1=\frac{1}{\ell}\D x^1, \quad \D\xi^2=\ell \,\D x^2, 
 \quad \ell=\sqrt{\frac{1+a}{1-a}}.  
\label{eq:x12}
\end{equation}
This relation can be used in lieu of Eq.~\eqref{eq:gg} when the coordinate axes $x^\alpha$, $\xi^\alpha$ are oriented along and normal to the director. 

%
%
A simple example is a \emph{vortex} defect, leading to an axisymmetric conical surface. Using the cylindrical coordinates $x^1=r,\, x^2=\phi,\, x^3=z$ with the Euclidean metric $g_{11}=g_{33}=1, \; g_{22}=r^2$, we obtain for a target surface with the elevation $z=h(r)$ and local radius $\rho(r)$ the diagonal metric tensor with the elements
\begin{equation}
\gamma_{11}= h_{,r}^2+\rho_{,r}^2, \qquad \gamma_{22}=\rho^2,
  \label{eq:g}
\end{equation}
The functions $\rho(r)$ and $h(r)$ are bound by the incompressibility constraint
\begin{equation}
(h_{,r}^2+\rho_{,r}^2)\rho^2 =r^2,
  \label{eq:h}
\end{equation}
which is integrated to
\begin{equation}
h(r) = \int \left( \frac{r^2}{\rho^2} - \rho_{,r}^2 \right)^{1/2} \D r.
  \label{eq:hr}
\end{equation}
The elements of the normalized tensor order parameter that corresponds to the director aligned everywhere along the angular coordinate are $Q^1{}_1 =-1, \; Q^2{}_2= 1, \; Q^1{}_2=0$. Then Eq.~\eqref{eq:gg} yields $\gamma_{11}= \ell^2, \; \gamma_{22}= (r/\ell)^2$. Comparing the latter relation with Eq.~\eqref{eq:g} directly defines $\rho(r)$, while $h(r)$ can be computed using Eq.~\eqref{eq:hr}.

The simplest solution, obtained at $a=$ const, is a straight cone $\rho = r/\ell , \;  h=r \sqrt{\ell^2- \ell^{-2}}$ with the Gaussian curvature vanishing everywhere except the singularity at the apex. 
The conical angle is approximated by $\pi/2 -\arctan (\rho/h) \approx 2\sqrt{a} \approx  2\sqrt{\ell-1} $ at $a \ll 1$, which yields substantial changes even at extensions of few percentage points, \emph{e.g.}\ the angle about 70$^\circ$ at $a \approx \ell-1 \approx 0.03$.  

\begin{figure}[t]
\centering \begin{tabular}{cc} (a) & (b) \\
\includegraphics[width=.24\textwidth]{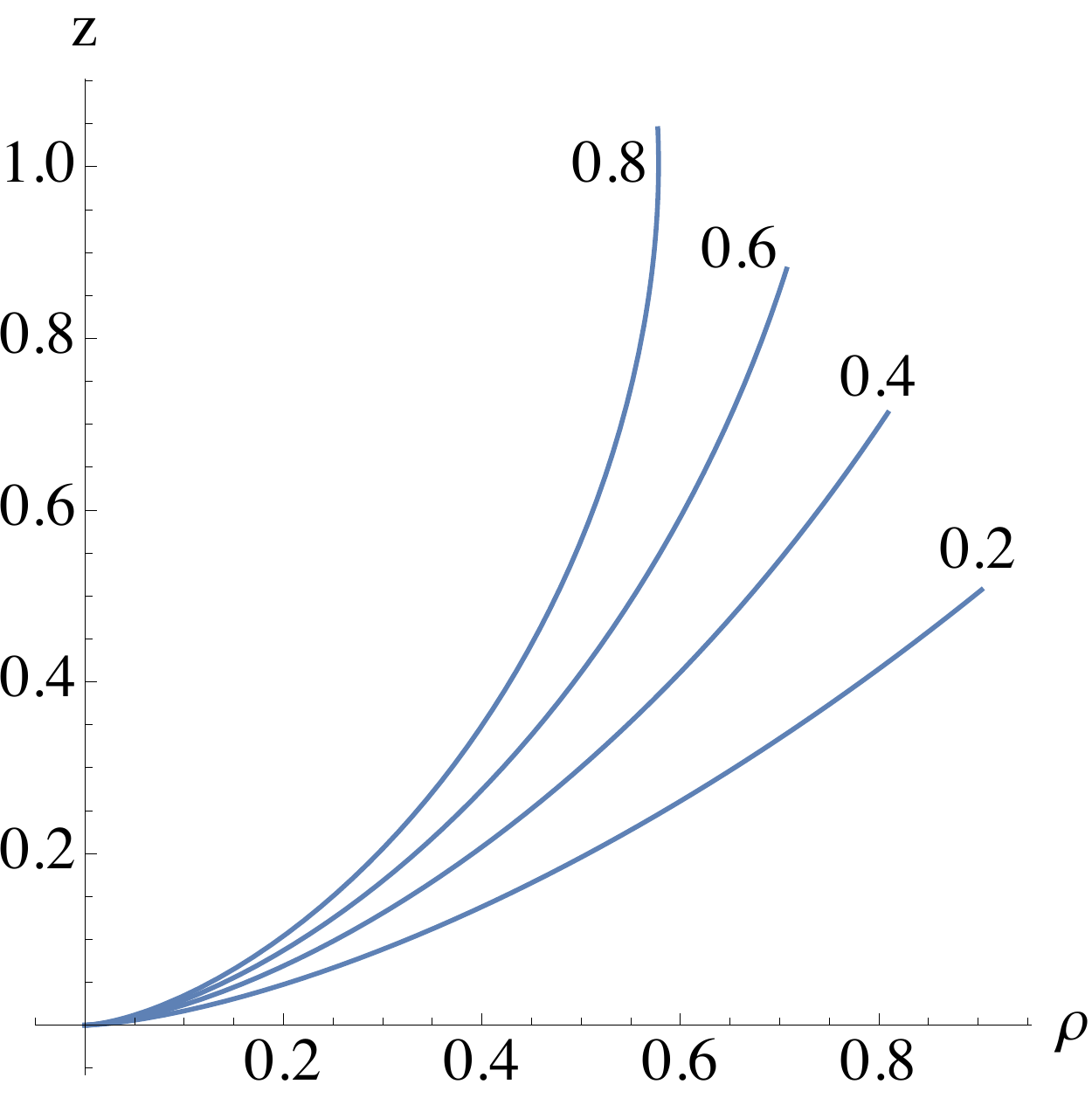}&
\includegraphics[width=.24\textwidth]{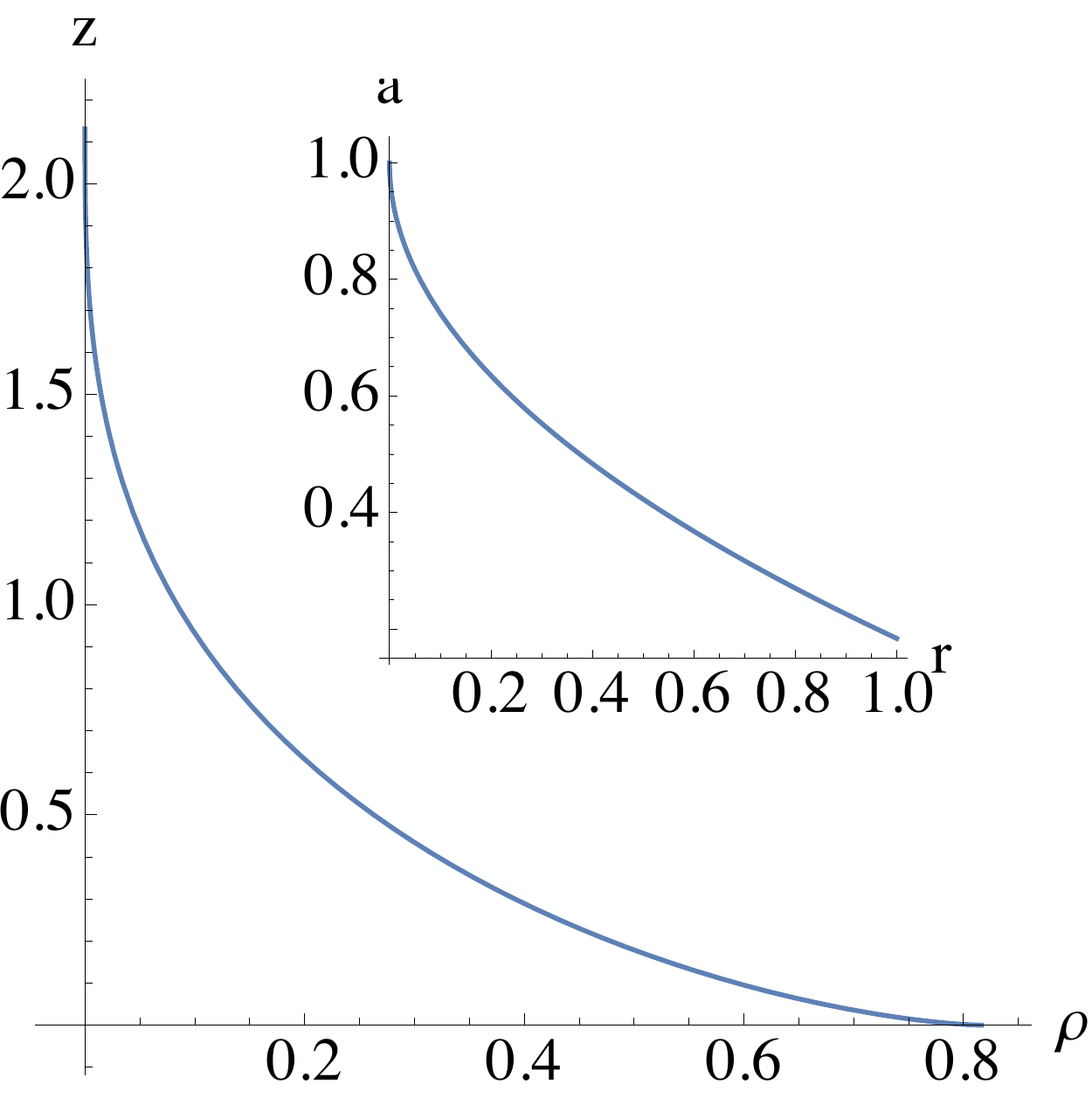}
\end{tabular}
\caption{(a) Bending the unit circle with $a(r)=kr$. Numbers at the curves show the values of $k$. (b) A singular shape obtained for $\rho(r)= n^{-1/2} r^n, \; n=3/2$ and the corresponding function $a(r)$}\label{axisym}
\end{figure}

Cones with curved generatrices and non-zero Gaussian curvature can be constructed by doping the nematic to assign a variable extension coefficient $\ell(r)$. Of particular interest are functions that vanish at the origin, thereby avoiding the topological singularity. If $a(r)=k r^n, \: n>0$ near the origin, the shape is approximated there as 
\begin{equation}
\rho(r) \approx r(1- k r^n) , \quad  h(r) = r^{n/2+1} \sqrt{\frac{8k}{n+2}},
  \label{eq:conern}
\end{equation}
which yields smooth shapes at the apex, \emph{e.g.} for $n=1$, $h \approx \rho^{3/2}\sqrt{8k/3}$ and for $n=2$, $h \approx \rho^{2}\sqrt{2k}$.
Some shapes, obtained from the unit circle by setting $a(r)=kr$ are shown in Fig.~\ref{axisym}(a). 

Alternatively, one can set an inverse problem by assigning a function $\rho(r)$ and calculating 
the requisite functions 
\begin{equation}
\ell(r)=\frac{r}{\rho} ,  \quad  a(r) =\frac{r^2-\rho^2}{r^2+\rho^2},
  \label{eq:cone1}
\end{equation}
with $h(r)$ defined by Eq.~\eqref{eq:hr}. Admissible functions $\rho(r)$ are restricted by the conditions $\rho < r, \; \rho|\rho_{,r}| \ge r$ following from Eqs.~\eqref{eq:h}, \eqref{eq:hr}, \emph{e.g.} power functions $\rho(r) =k r^n$ are admissible near the origin at $ 1<n < 2$. These  functions correspond to cones with concave generatrices and a sharper singularity at the apex, $h \propto \rho^{2/n-1}$ at $\rho \to 0$; an example is shown in Fig.~\ref{axisym}(b), 

%

%
\begin{figure}[b]
\centering \begin{tabular}{c} (a) \vspace{-4mm} \\
\includegraphics[width=.4\textwidth]{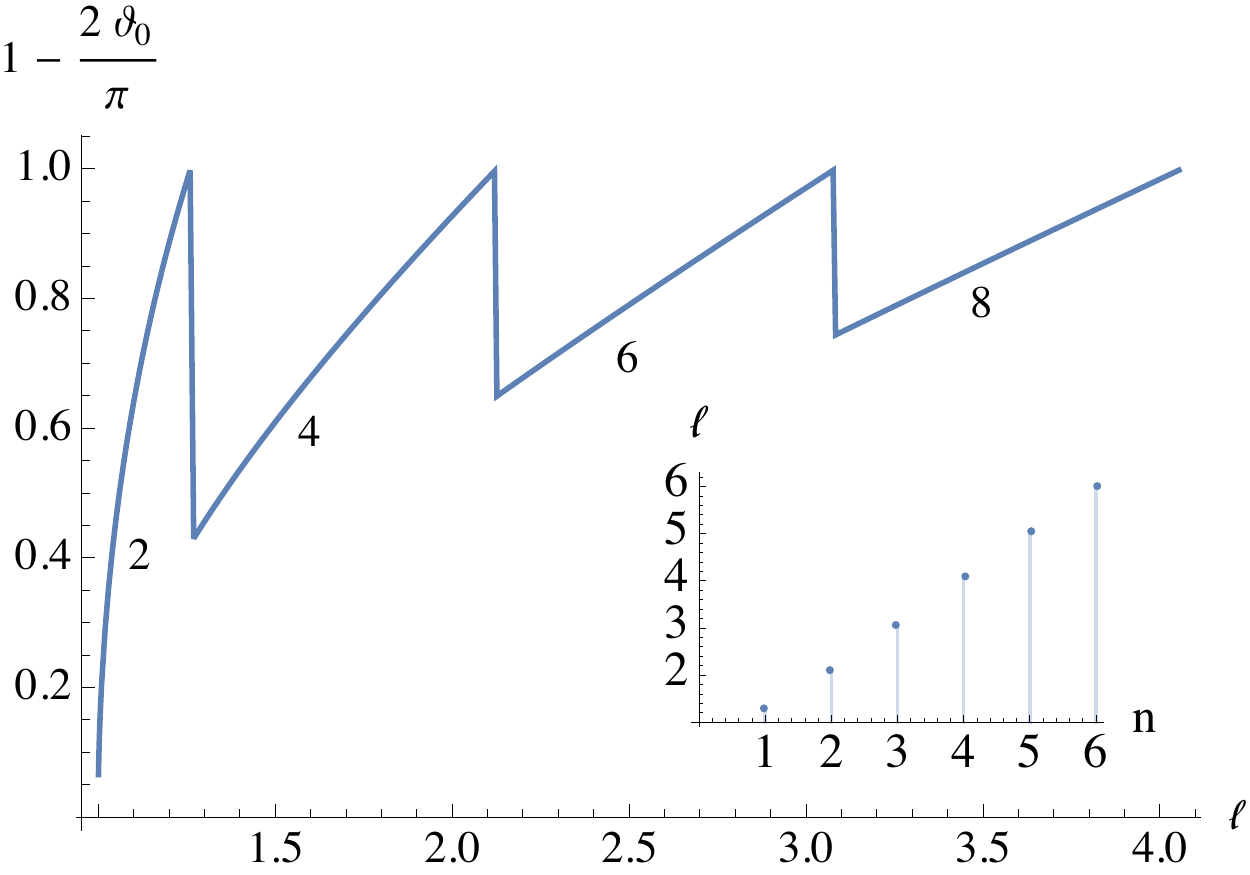} \vspace{3mm}\\
 (b) \vspace{-4mm} \\ \includegraphics[width=.4\textwidth]{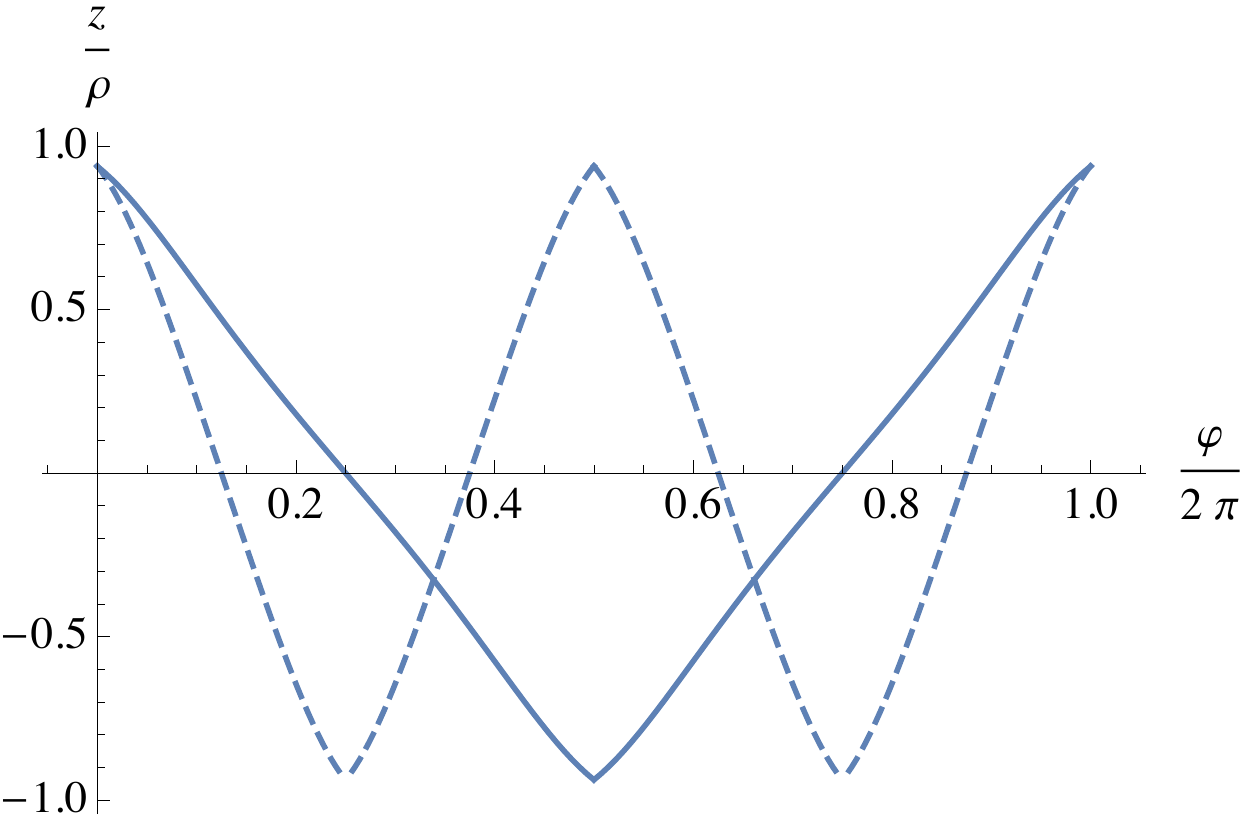}\vspace{3mm}\\
 (c) \vspace{-.1mm} \\ \includegraphics[width=.25\textwidth]{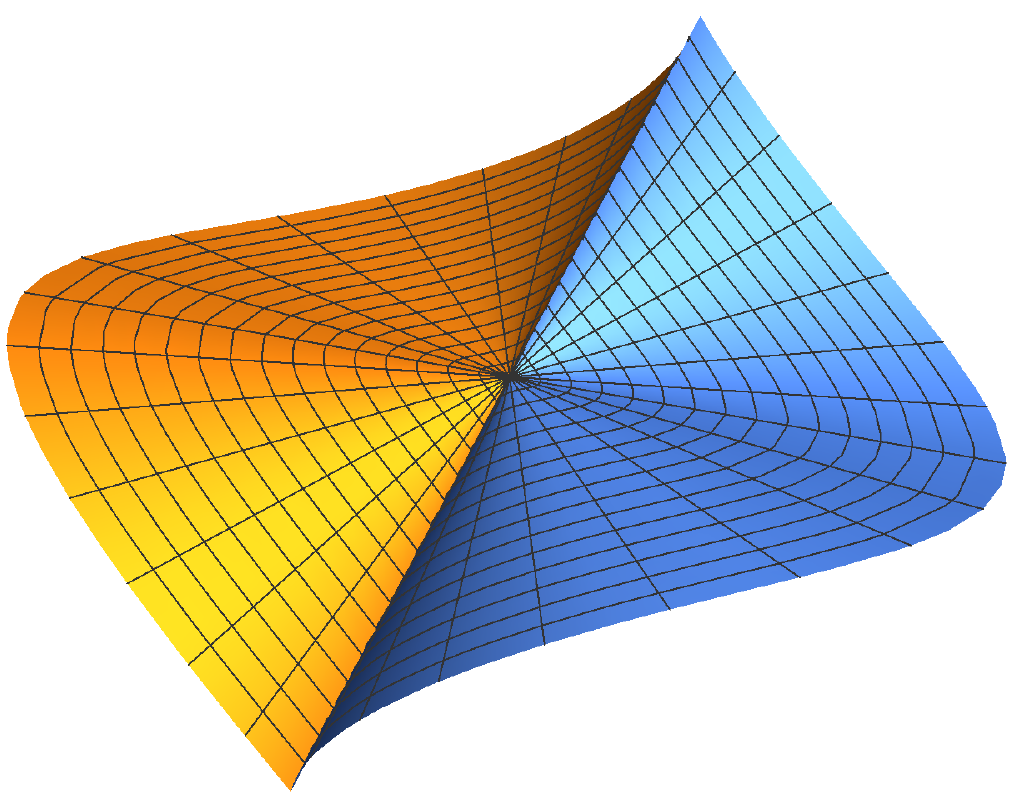}
 \includegraphics[width=.2\textwidth]{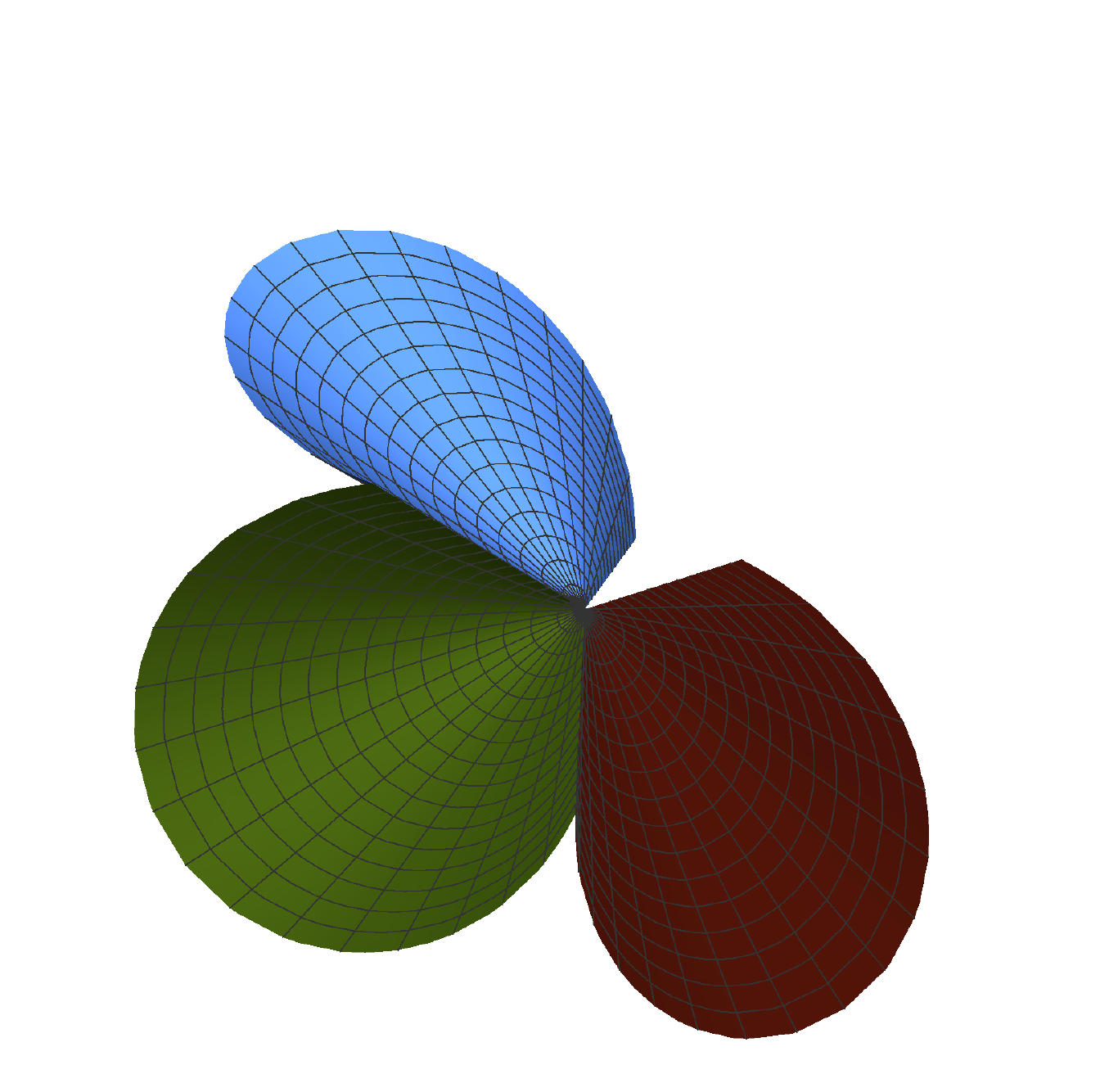}
\end{tabular}
\caption{(a) The polar angle at creases as a function of the extension coefficient $\ell$. Numbers at the curves denote the number of creases $2n$. Inset: the maximum values of $\ell$ sustaining $2n$ creases. (b) Relative elevation from the equator $z/\rho= \cos \vartheta$ as a function of the azimuthal angle $\varphi$. Solid line: $\ell=1.15, \; n=1$; dashed line: $\ell=1.75, \; n=2$. (c) 3D view of the anticones at the same values of $\ell$, showing the images of circles and radii in the original disk; one petal of the anticone with 4 creases is omitted for graphical clarity.}
\label{acone}
\end{figure}
%

The opposite case of radial contraction and angular expansion in the field of a charge one \emph{aster} defect leads to an asymmetric \emph{anticone} structure \cite{Warner12}. As the radii are contracted, the problem is equivalent to that of crumpling a disk to fit it into a sphere of a smaller diameter \cite{cerda,witten}; the resulting structures are known as $d$-cones. It is advantageous to compare the formation of cone and anticone structures at $a=$ const. In both cases,  the director is oriented along the coordinate lines, and therefore the extension and contraction should remain constant on the images of both circles and radii in the original flat disk. The transformation is revealed most clearly in spherical geometry, as the radial lines are mapped in the bent shell onto spherical radii converging at the apex, while the points of the original circle are shifted along meridians on the spherical surface \cite{Warner12}. 

It is appropriate therefore to work in spherical rather than cylindrical coordinates. We set $x^1=\rho, x^2=\varphi, x^3=\vartheta$ with the 3D Euclidean metric $g_{11}=1, \; g_{22}=\rho^2 \sin^2\vartheta, \; g_{33}=\rho^2$ and the original flat disk placed at $\vartheta=\pi/2$. The  images of the radii in the disk are aligned with the spherical radii in both cones and anticones, and in both cases the director remains aligned with the images of, respectively, circles and radii. In the conical transformation, as each concentric circle on a flat disk shrinks uniformly, its image should move toward either pole with $\vartheta(r)$ remaining independent of the azimuthal angle, while the radial coordinate extends accordingly. When, on the opposite, the radii are shrunk while the circles extended, the images of planar circles must deform to oscillating curves with a variable $\vartheta(r,\varphi)$ to accommodate a circumference to radius length ratio exceeding $2\pi$ but in both cases the shifts are meridional and $\vartheta$ is the only variable changing as the disk is reshaped following NIT. 

Modes \emph{et al} \cite{Warner12} postulated that the vertical deflection $h =\cos\vartheta$ can be expressed as $n$th harmonic of the polar angle in cylindrical coordinates $r,\,\phi$. Their expression for the image $h(\phi),\,r(\phi)$ of a circle with the radius $\rho$ in the planar sheet reads
\begin{equation}
 h(\phi) =A r(\phi) \sin n\phi, \quad 
 r(\phi) =\rho \left(1+A^2 n^2 \sin^2 n\phi \right)^{-1/2}.
   \notag 
\end{equation}
Even though the integral length of  the image curve can be adjusted to $2\pi \ell \rho$, the extension along this curve, 
\begin{equation}
\widetilde{\ell}(\phi) = \left(h_{,\phi}^2+r_{,\phi}^2 + r^2\right)^{1/2},
  \label{eq:ga3}
\end{equation}
varies with $\phi$, contrary to the persistent normal orientation of the director. As the radial contraction is constant, incompressibility is violated locally as well.

We proceed, with the help of  the metric machinery, to derive an alternative solution satisfying the required constraints. The induced metric on a surface with the polar angle $\vartheta(\varphi)$, parametrized by $\rho, \varphi$, is diagonal with the elements
\begin{equation}
\gamma_{11}= 1, \qquad 
\gamma_{22}=\rho^2 \left(\sin^2\vartheta(\varphi)+ (\D\vartheta/\D\varphi)^2 \right).
  \label{eq:ga}
\end{equation}
The incompressibility constraint imposes the relation with the radial coordinate $r(\rho,\varphi)$ of a point on the flat disk that projects onto the point on the bent shell with the specified coordinates:  
\begin{equation}
r(\rho,\varphi)=\rho\sqrt{\sin^2\vartheta(\varphi)+ (\D\vartheta/\D\varphi)^2}.
  \label{eq:ha}
\end{equation}
As the spherical radius contacts uniformly by the factor $\ell$, we can set $r=\rho\ell$. Accordingly, Eq.~\eqref{eq:x12} becomes 
\begin{equation}
 \D \rho =\ell^{-1} \D r, \quad
 r \, \D\varphi \sqrt{\sin^2\vartheta +  (\D\vartheta/\D\varphi)^2} = r\,\D\varphi.
  \label{eq:phph}
\end{equation}
Both Eqs.~\eqref{eq:ha} and \eqref{eq:phph} lead to the same differential equation of $\vartheta(\varphi)$:
\begin{equation}
 \D\vartheta/\D\varphi =\pm \sqrt{\ell^{2}-\sin^2\vartheta}.
  \label{eq:tph}
\end{equation}
It is integrated to the relation defining the interval $\delta\varphi$ wherein $\vartheta$ deviates from the equator $\vartheta=\pi/2$ by a certain decrement:
\begin{equation}
\delta\varphi(\vartheta) = \ell^{-1}\left[ K(\ell^{-1}) -F(\vartheta,\ell^{-1})\right], 
  \label{eq:tphs}
\end{equation}
where $F(\vartheta,x)$ is the elliptic integral of the first kind and $K(x)=F(\pi/2,x)$ is the complete elliptic integral.

Since, according to Eq.~\eqref{eq:tph}, $\D\vartheta/\D\varphi$ does not vanish anywhere, no smooth solutions satisfying $\vartheta(\varphi)=\vartheta(\varphi+2\pi)$ exist. The anticone can be, however, assembled from $2n$ patches separated by creases at $\phi=\phi_k$ where $\D\vartheta/\D\phi$ changes sign; the creases go all the way along spherical radii to the apex. To ensure continuity and periodicity, $\phi_k = \pi k/n, \; k=1, \ldots, 2n$ should be spaced at equal intervals. The value of  $ |\pi/2-\vartheta_0|$ at the creases is defined by the condition  $2n\delta\varphi (\vartheta_0)=\pi $. The minimal number of creases, attained when the creases converge at the poles $\vartheta=0$ and $\vartheta=\pi$, is  
\begin{equation}
2n= \pi  / \delta\varphi (0)= \pi  \ell / K(\ell^{-1}) .
  \label{eq:tphs1}
\end{equation}
This number grows with $\ell$ as seen in Fig.~\ref{acone}(a). Examples of 2D and 3D representations of anticones at selected values of $\ell$ are presented in Fig.~\ref{acone}(b,c). 
 
The creases are expected to smooth out with the curvature radius of the order of shell thickness. The minimal number of creases should be selected to minimize bending energy. Unlike crumpling of isotropic sheets where it may be energetically favorable to develop several conical singularities \cite{witten}, a symmetric set of  creases going all the way to the boundary cannot be avoided here in view of the constraint on a constant extension of images of the circles surrounding the aster singularity. The latter is, however, apt to split in a sufficiently large nematic disk into a pair of half-charged defects. In this case, a shell with two conical singularities should form following NIT. The geometry of extension and contraction is, however, substantially different in this case, and will be the subject of a further numerical study.
As before, the theory can be extended to extension coefficients $\ell(r)$ varying with radius due to doping. In this case, the number of creases would change abruptly at radial positions corresponding to critical values of $\ell$ defined by Eq.~\eqref{eq:tphs1} and shown in the inset of Fig.~\ref{acone}(a).

The scalar order parameter is naturally variable even without doping, as it must vanish at the defect location \cite{book}. As the metric factor $a$ goes down to zero accordingly, this provides an effective way to resolve the singularity at the apex. Modes \emph{et al} \cite{Warner12} resolved the curvature singularity in another way, by taking into account a finite bending energy of the shell and approximating the tip of a cone by a sphere with a radius minimizing the sum of bend and stretch energies. The effective curvature radius must be in this case of the same order of magnitude as the film thickness, and, if the latter is smaller than the healing length quantifying distortions of the nematic order parameter, its vanishing at the singularity would lead to a larger curvature radius without imposing any stretches. 

The most common way of description of non-uniform nematic fields \cite{degennes}, using the unit orientation vector $n^\alpha$, is restricted to perfectly oriented uniaxial media and is not applicable near defects. A far more complicated expression for the distortion energy, including the gradients of both director and scalar order parameter, was suggested by Ericksen  \cite{Ericksen}. A most general quadratic expression in terms of the tensor order parameter would contain a rank six elastic tensor. Since the tensor $\textbf{Q}$ has in 2D only two independent components and the energy should contain only squares of the two distinct operators $n^\alpha \partial_\alpha$ and $n^\alpha \epsilon_\alpha{}^\beta\partial_\beta$ (where $\epsilon_{\alpha\beta}$ is the Levi-Civita antisymmetric tensor), the number of independent coefficients reduces from $2^6$ to six, as compared to Ericksen's eight coefficients in 3D. Even this number is, however, uncomfortably large. The alternative is an expression neglecting the elastic anisotropy with respect to the director \cite{sheng,epj13}. In this formulation, the normalized 2D Landau--de Gennes Lagrangian reads
\begin{align}
{\cal L} = & -\frac{1}{4} Q_{\alpha\beta}Q^{\alpha\beta} +
\frac{1}{16}\left(Q_{\alpha\beta}Q^{\alpha\beta} \right)^2 \notag \\
& + \frac{\kappa_1}{2}  g^{\mu\nu} Q^{\alpha}{}_{\mu;\alpha} Q^\beta{}_{\nu;\beta}
+ \frac{\kappa_2}{4} g^{\mu\nu} Q^\alpha{}_{\beta ;\mu} Q^\beta{}_{\alpha;\nu}.  
\label{eq:LQdef}
\end{align}
This expression, containing only two independent elastic coefficients in either 2D or 3D, does not distinguish between energies of distortions parallel and normal to the director; in a perfectly aligned 2D material it is equivalent to the frequently used one-constant approximation that does not discriminate between splay and bend energies. We will further use this more convenient expression, in spite of these limitations. 

\begin{figure}[b]
 \centering 
\includegraphics[width=.4\textwidth]{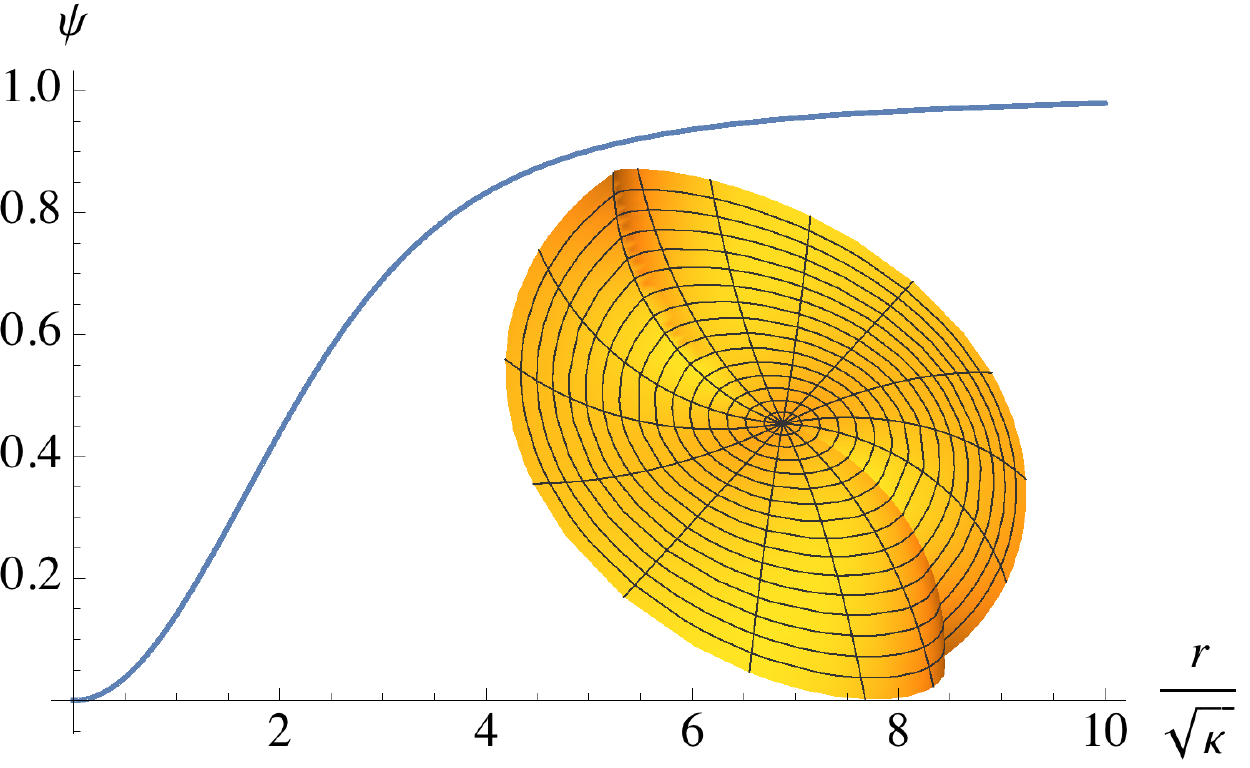}
\caption{\label{tip}The function $\psi(r)$. Inset: the shape of the anticone near the apex; $a_0=1/2, \; \rho<\sqrt{\kappa}$.}
\end{figure}

The Lagrangian \eqref{eq:LQdef} can be rewritten in Cartesian coordinates $x,y$ in terms of the functions $p$, $q$ in Eq.~\eqref{eq:pq} as \cite{epj13}
\begin{align}
  {\cal L} &=  - \frac{1}{2}\left(q^2 + p^2\right)+
 \frac{1}{4}\left(q^2 +p^2\right)^2  \notag \\
& +  \frac{\kappa_1}{2}\left[\left( p_{,x} + q_{,y} \right)^2 
+ \left( q_{,x} - p_{,y} \right)^2 \right]
\notag \\ &
 + \frac{\kappa_2}{2}\left[ p_{,x}^2 +
p_{,y}^2 + q_{,x}^2 + q_{,y}^2 \right]. 
 \label{eq:pqxy} \end{align}
The equilibrium equations obtained by varying Eq.~\eqref{eq:pqxy} reduce \cite{def} to a single equation for the complex variable $\chi=p+\I q =\psi\,\E^{2\I \theta}, \; \psi=|\chi|=\sqrt{p^2+q^2}$:
\begin{equation}
 \kappa \nabla^2 \chi + \chi - |\chi|^2 \chi =0,
\label{eq:evolch}
\end{equation}
where $\nabla^2$ is the Laplacian and $\kappa=\kappa_1+\kappa_2 \ll 1$ is the squared healing length. With the adopted normalization, the metric factor is $a=a_0\psi$, where $a_0$ is its asymptotic constant value outside the defect core, while $\theta=\pm\phi$ or $\phi\pm\pi/2$, respectively, for the charge one aster or vortex singularity. In the adopted approximation, there is no distinction between the both, so that Eq.~\eqref{eq:evolch} is equally applicable to cones and anticones. The resulting equation of $\psi$ is identical to the equation of dissipative dynamics of a vortex of double charge in a complex scalar field, and is written in polar coordinates as  
\begin{equation}
 \frac{\kappa}{r} \frac{\D}{\D r}\left(r\frac{\D\psi}{\D r} \right) 
  + \psi \left(1 - \frac{4\kappa}{r^2} -\psi^2 \right) =0.
\label{eq:evolps}
\end{equation}
The numerical solution is plotted in Fig.~\ref{tip}. 

Near the origin, $\psi \approx 0.153\, r^2 /\kappa$, and, according to Eq.~\eqref{eq:conern}, the apex of a cone is smoothed out at $O(\sqrt{\kappa})$ distances to a parabolic cap $z(\rho) \approx 0.39 \sqrt{a_0/\kappa}\, \rho^2$. 
In anticones, creases in excess of two must terminate within the healing length as $a \to 0$ at $\rho \to 0$. In the remaining two-crease structure, the maximum deviation of the polar angle from the equator is well approximated as $|\pi/2 -\vartheta_0| \approx 3.25 \sqrt{a}\approx 1.27 \sqrt{a_0/\kappa}\,\rho$ at $a \ll 1$. Therefore the maximum elevation $z_0 =\rho\cos \vartheta_0 \approx 1.27 \sqrt{a_0/\kappa}\,\rho^2$ is of the same order of magnitude as for a cone. The 3D shape near the apex is shown in the inset of Fig.~\ref{tip}. 

The two ways of resolving the tip singularities are complementary. By the estimates in Ref.~\cite{Warner12}, stretching at the tip is the preferred mechanism when the nematic healing length is in the nanometer range. The latter, however, grows indefinitely as the NIT point is approached, so that the decay of the order parameter should become the prevailing mechanism sufficiently close to NIT.    

\emph{Acknowledgment}. This research is supported by Israel Science Foundation (grant 669/14).


\end{document}